\documentclass[conference,10pt]{IEEEtran}
\usepackage{graphicx}
\usepackage{amssymb}
\usepackage{amsmath}

\DeclareMathOperator{\vect2}{vec}
\DeclareMathOperator{\diag}{diag}

\usepackage{algorithmic}
\usepackage{array}
\usepackage[caption=false,font=footnotesize]{subfig}
\usepackage{stfloats}
\usepackage{url}
\usepackage[short]{optidef}
\usepackage{bm}
\usepackage{makecell}
\usepackage{algorithmic}
\usepackage{booktabs}
\usepackage{adjustbox}
\usepackage{multirow}
\usepackage{cite}
\usepackage[linesnumbered,ruled,vlined]{algorithm2e}
\usepackage{setspace}
\def\BibTeX{{\rm B\kern-.05em{\sc i\kern-.025em b}\kern-.08em
    T\kern-.1667em\lower.7ex\hbox{E}\kern-.125emX}}

\SetKwInput{KwInput}{Input}                
\SetKwInput{KwOutput}{Output}              
\ifCLASSINFOpdf
\else   
\fi

\begin{document}
\title{A Rate-Splitting Strategy to Enable Joint Radar Sensing and Communication with Partial CSIT}

\author{
        \IEEEauthorblockN{Rafael Cerna-Loli, Onur Dizdar and Bruno Clerckx}
        \IEEEauthorblockA{Department of Electrical and Electronic Engineering \\
        Imperial College London, London, U.K.
        \\
        Email: \{rafael.cerna-loli19, o.dizdar, b.clerckx\}@imperial.ac.uk}
        }

\maketitle

\begin{abstract}
In order to manage the increasing interference between radar and communication systems, joint radar and communication (RadCom) systems have attracted increased attention in recent years, with the studies so far considering the assumption of perfect Channel State Information at the Transmitter (CSIT). However, such an assumption is unrealistic and neglects the inevitable CSIT errors that need to be considered to fully exploit the multi-antenna processing and interference management capabilities of a joint RadCom system. In this work, a joint RadCom system is designed which marries the capabilities of a Multiple-Input Multiple-Output (MIMO) radar with Rate-Splitting Multiple Access (RSMA), a powerful downlink communications scheme based on linearly precoded Rate-Splitting (RS) to partially decode multi-user interference (MUI) and partially treat it as noise. In this way, the RadCom precoders are optimized in the presence of partial CSIT to simultaneously maximize the Average Weighted Sum-Rate (AWSR) under QoS rate  constraints and minimize the RadCom Beampattern Squared Error (BSE) against an ideal MIMO radar beampattern. Simulation results demonstrate that RSMA provides the RadCom with more robustness, flexibility and user rate fairness compared to the baseline joint RadCom system based on Space Division Multiple Access (SDMA).
\end{abstract}


\IEEEpeerreviewmaketitle

\section{Introduction}
Radar systems are vital in public safety and military applications, where it is necessary to identify relevant targets with a high resolution estimation of their associated angle, range and velocity. This requires that radar systems are allocated sufficient electromagnetic (EM) spectrum resources to collect all the necessary information with a single radar pulse \cite{mit_radar}. On the other hand, next generation wireless communication systems, such as the 5G-New Radio (NR) mobile communication networks and Internet of Things (IoT), also demand large spectrum resources to offer high data rate services with a guaranteed Quality-of-Service (QoS) level. Due to insufficient available bandwidth, specially in sub-10 GHz bands, increased spectrum congestion and inter-system interference is expected without careful simultaneous deployment planning of radar and communication systems \cite{ucl_crss}. This spectrum congestion issue is the focus of Communication and Radar Spectrum Sharing (CRSS) research \cite{survey_pu_su}, where different studies have recently been made in order to optimize different performance metrics of spectrum sharing radar and communication (RadCom) systems by employing techniques such as interference mitigation, beamforming, and optimum waveform design. Nevertheless, these efforts can generally be classified into two categories: coexistent RadComs and joint RadComs.

Coexistent RadCom design considers that the radar and communication parts are deployed separately, with independent hardware and signal processing units, but share substantial information between each other in order to optimize their individual performance \cite{coop_1}. To achieve this, the RadCom may employ a control center or mediator to relay the necessary information and keep them synchronized. Although theoretically functional, including this external element would greatly increase hardware costs and required computational power. This issue is bypassed with a joint RadCom design as radar and communication modules are deployed with unified hardware and signal processing units \cite{radcom_survey}. Thus, this approach is also the most suitable for a long-term development of wireless systems and EM spectrum allocation. Advantages of a joint design also include highly-directional beamforming, minimum delay, enhanced security and privacy, and dynamic computational resource allocation.

This paper follows our earlier work in \cite{chengcheng} and extends it to optimize the precoders of a joint RadCom system in the more realistic and important partial CSIT setting for the first time. In order to achieve this, a Rate-Splitting Multiple Access (RSMA) communications module is considered to operate jointly with a Multiple-Input Multiple-Output (MIMO) radar module. As it will be demonstrated in the following sections, RSMA constitutes a robust interference management framework in the presence of CSIT errors that aims to mitigate multi-user interference (MUI) by splitting the user data streams into common streams decoded by all users (partially decoding MUI), and private streams decoded only by its intended user (partially treating MUI as noise) \cite{rsma_lina}. In the context of a joint RadCom design with partial CSIT, RSMA offers a special advantage as the beampattern of the common stream can be used to approximate a highly-directional transmit beampattern, which greatly increases the detection capabilities of the MIMO radar module, while also providing flexibility to comply with QoS rate constraints. 

\section{Joint RadCom System Model}
\label{system_model_section}
\begin{figure}[t!]
		\centering
        \includegraphics[scale = 0.53]{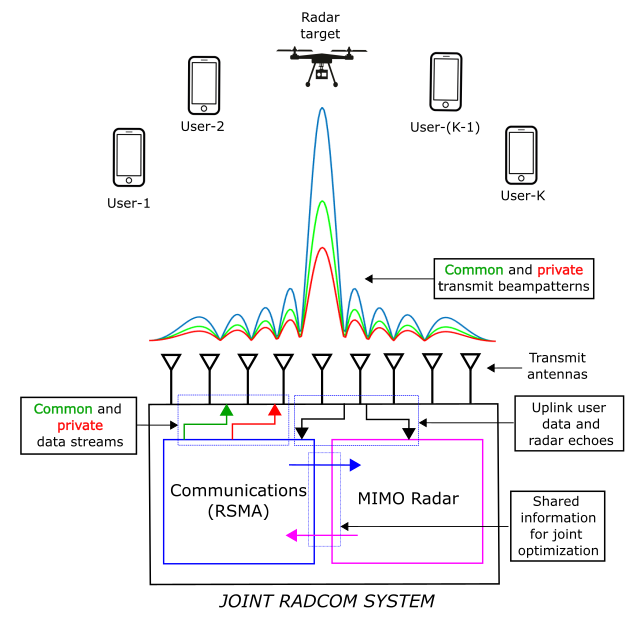}
		\caption{Proposed joint RadCom system model.}
		\label{fig:radcom_system_model}
\end{figure}

Consider a joint RadCom, with a uniform linear array of $N_t$ transmit antennas and a total available transmit power $P_t$, that serves $K$ single antenna communication users, indexed by the set $\mathcal{K} = \{1,\dots,K\}$, and tracks a single radar target as depicted in Fig. \ref{fig:radcom_system_model}. It employs an RSMA communications module and a mono-static MIMO radar module that share information, such as transmit communication signals and radar target parameters, to perform joint precoder optimization.

\subsection{RSMA-RadCom Signal Model}
In this subsection, the operation of the RSMA communications module is described. The intended message for user-$k$ $W_k$ is split into a common part $W_{c,k}$ and a private part $W_{p,k}$. Then, the common parts of all $K$ users $\{W_{c,1},\dots,W_{c,K}\}$ are encoded into a single common stream $s_c$, while the private parts $\{W_{p,1},\dots,W_{p,K}\}$ are encoded into $K$ different private streams $\{s_1,\dots,s_K\}$. The data stream vector $\mathbf{s} = [s_c,s_1,\dots,s_K]^T \in \mathbb{C}^{(K+1)\times 1}$ is linearly precoded using the precoder $\mathbf{P} = [\mathbf{p}_c,\mathbf{p}_1,\dots,\mathbf{p}_K] \in \mathbb{C}^{N_t \times (K+1)}$, where $\mathbf{p}_c$ is the common stream precoder and $\mathbf{p}_k$ is the private stream precoder for user-$k$. The transmitted signal $\mathbf{x} \in \mathbb{C}^{N_t \times 1}$ is then given by
\begin{equation}
    \mathbf{x} = \mathbf{P}\mathbf{s} = \mathbf{p}_cs_c + \sum_{k=1}^{K}\mathbf{p}_ks_k.
    \label{rsma_transmit_signal}
\end{equation}
It is proposed that the communication signal in (\ref{rsma_transmit_signal}) is also used for MIMO radar purposes following the work in \cite{friedlander_radar}. It is shown in \cite{friedlander_radar} that the optimum design of the transmit signal covariance matrix $\mathbf{R}_{\mathbf{x}}$ of a MIMO radar can be achieved in a simplified manner by generating the transmitted signal $\mathbf{x}$ as a linear combination of independent signals, which effectively matches the signal model in (\ref{rsma_transmit_signal}). In this way, optimization of $\mathbf{R}_{\mathbf{x}}$ is reduced to optimization of the precoder matrix $\mathbf{P}$. 

The  signal received by user-$k$ is then given by
\begin{equation}
    \begin{split}
        y_k &= \mathbf{h}_k^H\mathbf{P}\mathbf{s} + n_k \\
        &= \mathbf{h}_k^H\mathbf{p}_c s_c+ \mathbf{h}_k^H\mathbf{p}_k s_k+\overbrace{\sum_{j\neq k,j\in\mathcal{K}}\mathbf{h}_k^H\mathbf{p}_j s_j}^{\text{MUI}} + n_k,
    \end{split}
    \label{rsma_eq}
\end{equation}
where $\mathbf{h}_k \in \mathbb{C}^{N_t \times 1}$ is the channel between the RadCom and user-$k$, and $n_k\;\mathtt{\sim}\; \mathcal{CN}(0,\sigma_{n,k}^2)$ is the Additive White Gaussian Noise (AWGN) at user-$k$. Without loss of generality, it is assumed that $\sigma_{n,k}^2 = \sigma_n^2 = 1,\; \forall k \in \mathcal{K}$.

The Signal-to-Interference-and-Noise Ratio (SINR) of the common stream at user-$k$ is given by
\begin{equation}
    \gamma_{c,k} = \frac{|\mathbf{h}_k^H\mathbf{p}_c|^2}{\sum_{k\in\mathcal{K}}|\mathbf{h}_k^H\mathbf{p}_k|^2+\sigma_{n,k}^2}.
    \label{commonsinr}
\end{equation}
After decoding the common stream, Successive Interference Cancellation (SIC) is applied to remove the obtained estimation from the received signal $y_k$ and then decode the private stream. The SINR of the private stream at user-$k$ is then given by
\begin{equation}
    \gamma_k = \frac{|\mathbf{h}_k^H\mathbf{p}_k|^2}{\sum_{j\neq k,j\in\mathcal{K}}|\mathbf{h}_k^H\mathbf{p}_j|^2+\sigma_{n,k}^2}.
    \label{privatesinr}
\end{equation} 
Therefore, the achievable rate of the common stream for user-$k$ is  $R_{c,k}(\mathbf{P}) = \log_2(1+\gamma_{c,k})$ and the achievable rate of its corresponding private stream is $R_k(\mathbf{P}) = \log_2(1+\gamma_k)$. In order to ensure that all $K$ users are able to decode the common stream, it must be transmitted at a rate no larger than $R_c(\mathbf{P}) = \min\{R_{c,1}(\mathbf{P}),\dots,R_{c,K}(\mathbf{P})\}$, with the portion of the total common stream rate assigned to user-$k$ being given by $C_k$, such that $\sum_{k\in\mathcal{K}}C_k=R_c(\mathbf{P})$. 

\subsection{Channel State Information Model}
The Channel State Information (CSI) is modeled by $\mathbf{H}=\hat{\mathbf{H}}+\Tilde{\mathbf{H}}$, where $\mathbf{H} = [\mathbf{h}_1,\dots,\mathbf{h}_K]$ is the real channel, $\hat{\mathbf{H}} = [\hat{\mathbf{h}}_1,\dots,\hat{\mathbf{h}}_K]$ is the estimated channel at the RadCom, and $\Tilde{\mathbf{H}} = [\Tilde{\mathbf{h}}_1,\dots,\Tilde{\mathbf{h}}_K]$ is the estimation error matrix. It is also assumed that $\mathbf{h}_k$, $\hat{\mathbf{h}}_k$ and $\Tilde{\mathbf{h}}_k$ have i.i.d complex Gaussian entries drawn from the distributions $\mathcal{CN}(0,\sigma_k^2)$, $\mathcal{CN}(0,\sigma_k^2-\sigma_{e,k}^2)$ and $\mathcal{CN}(0,\sigma_{e,k}^2)$ respectively, for each $k\in\mathcal{K}$. The parameter $\sigma_{e,k}^2 \triangleq \sigma_k^2P_t^{-\alpha}$ is the CSIT error power for user-$k$, where $\alpha \in [0,\infty)$ is the CSIT quality scaling factor \cite{joudeh}. $\alpha \rightarrow \infty$ corresponds to perfect CSIT while $\alpha = 0$ represents partial CSIT with finite precision. In this work, perfect Channel State Information at the Receiver (CSIR) and partial CSIT are assumed, where the latter indicates that the RadCom only knows $\hat{\mathbf{H}}$ and the conditional CSIT error distribution $f_{\text{H}|\hat{\text{H}}}(\mathbf{H}|\hat{\mathbf{H}})$.

\normalsize
\section{Performance Metrics and Problem Formulation}
\label{problem_section}

In this section, the performance metrics for communications and radar sensing are introduced and used to define the joint RadCom optimization problem.

\subsection{Communications Metric: Average Weighted Sum-Rate}
To achieve maximum user rates, (\ref{commonsinr}) and (\ref{privatesinr}) need to be jointly maximized. However, computation of the exact precoders that maximize the common and private SINRs is not possible with partial CSIT. On one hand, a naive strategy would be to treat the estimated channel $\hat{\mathbf{H}}$ as perfect CSIT, which would result in increased MUI, inefficiency in the precoder power allocation and, ultimately, transmission at undecodable rates. On the other hand, a more resilient strategy is to adapt the precoder matrix $\hat{\mathbf{P}}=\mathbf{P}(P_t,\hat{\mathbf{H}})$ to send the common stream and the private streams at their Ergodic Rates (ERs), representations of the long-term rates over all channel states for the distribution $f_{\text{H}}(\mathbf{H})$. The ERs for user-$k$ are given by $\mathbb{E}_\text{H}\{R_{c,k}\}$ and  $\mathbb{E}_\text{H}\{R_{k}\}$ for the common and private stream respectively. Additionally, the common ER to guarantee successful decoding by all $K$ users is given by $\min_k\{\mathbb{E}_\text{H}\{R_{c,k}\}\}_{k=1}^K$.

Although the ERs cannot be directly maximized without perfect CSIT, optimization of the ERs under partial CSIT can be achieved by maximizing the average Rates (ARs), short-term measures of the expected performance over $f_{\text{H}|\hat{\text{H}}}(\mathbf{H}|\hat{\mathbf{H}})$, of the common and private streams for each channel estimate $\hat{\mathbf{H}}$. The ARs for user-$k$ are then given by $\bar{R}_{c,k}\triangleq\mathbb{E}_{\text{H}|\hat{\text{H}}}\{R_{c,k}|\hat{\mathbf{H}}\}$ and  $\bar{R}_k\triangleq\mathbb{E}_{\text{H}|\hat{\text{H}}}\{R_{k}|\hat{\mathbf{H}}\}$ for the common and private stream respectively. Additionally, the common AR for all users is given by $\bar{R}_c\triangleq\min_k\{\mathbb{E}_{\text{H}|\hat{\text{H}}}\{R_{c,k}|\hat{\mathbf{H}}\}\}_{k=1}^K$.  The Average Weighted Sum-Rate (AWSR) metric is then defined as
\begin{equation}
    \text{AWSR}(\hat{\mathbf{P}}) = \sum_{k\in\mathcal{K}}\mu_k(\bar{C}_k+\bar{R}_k(\hat{\mathbf{P}})),
    \label{awsr_def}
\end{equation}
where $\mu_k$ is the weight assigned to user-$k$.

\subsection{Radar Sensing Metric: Beampattern Squared Error}
As shown in \cite{friedlander_radar}, the detection capabilities of a MIMO radar can be improved by appropriately designing the covariance matrix $\mathbf{R}_{\mathbf{x}} \in \mathbb{C}^{N_t \times N_t}$ of the transmitted signal $\mathbf{x}$ to approximate a highly directional transmit beampattern $\bm{P}_d$. Thus, the radar sensing metric, the Beampattern Squared Error (BSE), can be defined as $ \sum_{m=1}^M|\alpha \bm{P}_d(\theta_m)-\mathbf{a}_t^H(\theta_m)\mathbf{R}_{\mathbf{x}}\mathbf{a}_t(\theta_m)|^2$, where $\alpha$ is the scaling factor of $\bm{P}_d$, $M$ is the total number of azimuth angle grids, $\theta_m$ is the $m^{th}$ azimuth angle grid,  $\mathbf{a}_t(\theta_m) = [1,e^{j2\pi\delta\sin(\theta_m},\dots,e^{j2\pi(N_t-1)\delta\sin(\theta_m)}]^T \in \mathbb{C}^{N_t \times 1}$ is the transmit antenna array steering vector at direction $\theta_m$ and $\delta$ is the normalized distance in units of wavelengths between antennas. In the context of the proposed RadCom transmission, the BSE is then given by
\begin{equation}
    \text{BSE}(\hat{\mathbf{P}}) = \sum_{m=1}^M|\alpha \bm{P}_d(\theta_m)-\mathbf{a}_t^H(\theta_m)\hat{\mathbf{P}}\hat{\mathbf{P}}^H\mathbf{a}_t(\theta_m)|^2,
\end{equation}
where $\bm{P}_t(\theta_m)=\mathbf{a}_t^H(\theta_m)\hat{\mathbf{P}}\hat{\mathbf{P}}^H\mathbf{a}_t(\theta_m) =  \mathbf{a}_t^H(\theta_m)\hat{\mathbf{p}}_c\hat{\mathbf{p}}_c^H\mathbf{a}_t(\theta_m) + \sum_{k=1}^K\mathbf{a}_t^H(\theta_m)\hat{\mathbf{p}}_k\hat{\mathbf{p}}_k^H\mathbf{a}_t(\theta_m)$ is the RadCom transmit beampattern gain at direction $\theta_m$, which is formed by the sum of the individual beampattern gains corresponding to the common and private data streams.

\subsection{Problem Formulation}
The RadCom optimization problem with partial CSIT can then be defined for a given channel estimate $\hat{\mathbf{H}}$ as follows:
\begin{mini!}|s|[2]
  {\alpha,\bar{\mathbf{c}},\hat{\mathbf{P}}}{\begin{aligned}[t]-\sum_{k\in\mathcal{K}}\mu_k(\bar{C}_k+\bar{R}_k(\hat{\mathbf{P}}))\qquad\qquad\qquad\qquad\;\;\;\\~\mathllap{+\lambda{\sum_{m=1}^M|\alpha \bm{P}_d(\theta_m)-\mathbf{a}^H(\theta_m)\big(\hat{\mathbf{P}}\hat{\mathbf{P}}^H\big)\mathbf{a}(\theta_m)|^2}},\end{aligned}}
  {\label{main_design_awsr}}{}
  \addConstraint{\sum_{k'\in\mathcal{K}}\bar{C}_{k'}}{\leq \bar{R}_{c,k}(\hat{\mathbf{P}}),\quad\forall k \in \mathcal{K}\label{eq:C1Main_design_awsr}}{}
  \addConstraint{\bar{\mathbf{c}}}{\geq 0\label{eq:C2Main_design_awsr}}
  \addConstraint{\diag(\hat{\mathbf{P}}\hat{\mathbf{P}}^H)}{=\frac{P_t\mathbf{1}}{N_t}\label{eq:C3Main_design_awsr}}
  \addConstraint{\alpha}{>0}
  \addConstraint{(\bar{C}_k + \bar{R}_k(\hat{\mathbf{P}}))}{\geq \bar{R}_k^{th}\;,\;\forall k\in\mathcal{K},\label{eq:C4Main_design_awsr}}
\end{mini!}
\normalsize   
where $\bar{\mathbf{c}} = [\bar{C}_1,\dots,\bar{C}_K]^T \in \mathbb{R}_+^{K \times 1}$ is the variable vector that contains the portions of the common stream AR, $\bar{R}_c(\hat{\mathbf{P}})$, allocated to the communication users, $\lambda$ is the regularization parameter to prioritize either communications (maximizing the AWSR) or radar sensing (minimizing the BSE), and $\bar{R}_k^{th}$ is the minimum average rate for user-$k$. Constraint (\ref{eq:C1Main_design_awsr}) ensures that $\bar{R}_c(\mathbf{P})$ is decodable by all $K$ users. Constraint (\ref{eq:C2Main_design_awsr}) forces the entries of $\bar{\mathbf{c}}$ to be positive for feasible partitioning of $\bar{R}_c(\mathbf{P})$. Also, constraint (\ref{eq:C3Main_design_awsr}) is introduced as an average power constraint at each transmit antenna to avoid saturation of transmit power amplifiers in a practical scenario. Finally, constraint (\ref{eq:C4Main_design_awsr}) is the optional QoS rate constraint to guarantee user rate fairness.

\section{Precoder Optimization with partial CSIT}
\label{optimization_section}
Based on the work presented in \cite{chengcheng}, it is proposed that the non-convex optimization problem in (\ref{main_design_awsr}) is solved in an alternating manner by employing the method of Alternating Direction Method of Multipliers (ADMM).

The new optimization variable $\mathbf{v} = [\alpha, \bar{\mathbf{c}}^T, \vect2(\hat{\mathbf{P}})^T]^T \in \mathbb{R}_{++} \times \mathbb{R}_+^{K \times 1} \times \mathbb{C}^{N_t(K+1) \times 1}$ is introduced to handle all optimization variables in (\ref{main_design_awsr}). Then, selection matrices are defined as $\mathbf{D}_p = [\mathbf{0}^{(K+1)N_t\times (K+1)},\mathbf{I}_{(K+1)N_t}]$, $\mathbf{D}_c = [\mathbf{0}^{N_t \times (K+1)}, \mathbf{I}_{N_t}, \mathbf{0}^{N_t \times KN_t}]$ and $\mathbf{D}_k = [\mathbf{0}^{N_t \times (K+1+kN_t)}, \mathbf{I}_{N_t}, \mathbf{0}^{N_t \times (K-k)N_t}]$ $\forall k \in \mathcal{K}$, and selection vectors $\mathbf{f}_k = [\mathbf{0}^{1 \times k}, 1, \mathbf{0}^{1 \times [(K+1)N_t+K-k]}]^T$ $\forall k \in \mathcal{K}$, which are used to extract $\bar{C}_k = \mathbf{f}_k^T \mathbf{v}$.

The user ARs and $\bar{R}_{k}(\hat{\mathbf{P}})$ are expressed as $R_{c,k}(\hat{\mathbf{P}})=\eta_{c,k}(\vect2(\hat{\mathbf{P}}))=\eta_{c,k}(\mathbf{D}_p\mathbf{v})$ and $R_k(\hat{\mathbf{P}})=\eta_k(\vect2(\hat{\mathbf{P}}))=\eta_k(\mathbf{D}_p\mathbf{v})$. Then,  (\ref{main_design_awsr}) is reformulated in an ADMM expression as follows:
\begin{mini}
  {\mathbf{v},\mathbf{u}}{f_c(\mathbf{v})+g_c(\mathbf{v})+f_r(\mathbf{u})+g_r(\mathbf{u})}{}{}
  \addConstraint{\mathbf{D}_p(\mathbf{v}-\mathbf{u})}{=0},
  \label{admm_design}
\end{mini}
where $\mathbf{u} \in \mathbb{R}_{++} \times \mathbb{R}_+^{K \times 1} \times \mathbb{C}^{N_t(K+1) \times 1}$ is a new optimization variable introduced to fit the ADMM optimization definition and it is initialized as $\mathbf{u} = \mathbf{v}$. The functions $f_c(\mathbf{v})$ and $f_r(\mathbf{u})$ are defined as $f_c(\mathbf{v})=-\sum_{k\in\mathcal{K}}\mu_k(\mathbf{f}_k\mathbf{v}+\eta_k\big(\mathbf{D}_p\mathbf{v}\big))$ and $f_r(\mathbf{u})=\lambda\sum_{m=1}^M|\alpha \bm{P}_d(\theta_m)-\mathbf{a}^H(\theta_m)\big(\mathbf{D}_c\mathbf{u}\mathbf{u}^H\mathbf{D}_c^H
+\sum_{k\in\mathcal{K}}\mathbf{D}_k\mathbf{u}\mathbf{u}^H\mathbf{D}_k^H\big)\mathbf{a}(\theta_m)|^2$. Also, $g_c(\mathbf{v})$ is the indicator function of the communication feasible set $\mathcal{C}=\Big\{\mathbf{v}\Big|\sum_{k\in\mathcal{K}}\mathbf{f}_k^T\mathbf{v}\leq\eta_{c,k}(\mathbf{D}_p\mathbf{v})\Big\}$, and $g_r(\mathbf{u})$ is the indicator function of the radar feasible set $    \mathcal{R}=\Big\{\mathbf{u}\Big|\diag\big(\mathbf{D}_c\mathbf{u}\mathbf{u}^H\mathbf{D}_c^H+\sum_{k\in\mathcal{K}}\mathbf{D}_k\mathbf{u}\mathbf{u}^H\mathbf{D}_k^H\big)=\frac{P_t\mathbf{1}}{N_t}\Big\}$.

Finally, (\ref{admm_design}) is solved in an iterative updating manner as follows:
\begin{equation}
    \begin{aligned}
        \mathbf{v}^{t+1}:=&\arg \min_{\mathbf{v}}\big(f_c(\mathbf{v}) + g_c(\mathbf{v})\\&+(\rho/2)||\mathbf{D}_p(\mathbf{v}-\mathbf{u}^t)+\mathbf{d}^t||_2^2\big),
    \end{aligned}
    \label{vupdate}
\end{equation}
\begin{equation}
    \begin{aligned}
       \mathbf{u}^{t+1}:=&\arg \min_{\mathbf{u}}\big(f(\mathbf{u}) + g_r( \mathbf{u})\\&+(\rho/2)||\mathbf{D}_p(\mathbf{v}^{t+1}-\mathbf{u})+\mathbf{d}^t||_2^2\big),
    \end{aligned}
    \label{uupdate}
\end{equation}
\begin{equation}
    \begin{aligned}
        \mathbf{d}^{t+1}:=&\mathbf{d}^t+\mathbf{D}_p(\mathbf{v}^{t+1}-\mathbf{u}^{t+1}), 
    \end{aligned}
    \label{dupdate}
\end{equation}
where $\mathbf{d} \in  \mathbb{C}^{N_t(K+1) \times 1}$ is the ADMM scaled dual variable and $\rho$ is the ADMM penalty parameter that controls the optimization convergence speed. The methods to perform the $\mathbf{v}$-update and the $\mathbf{u}$-update are explained next.

\subsection{AWSR Maximization Sub-problem}
The $\mathbf{v}$-update sub-problem in (\ref{vupdate}) is reformulated as follows:
\begin{mini}|s|[2]
  {\bar{\mathbf{c}},\hat{\mathbf{P}}}{-\sum_{k\in\mathcal{K}}\mu_k[\bar{C}_k+\bar{R}_k(\hat{\mathbf{P}})]+\frac{\rho}{2}||\vect2(\hat{\mathbf{P}})-\mathbf{D}_p\mathbf{u}^t+\mathbf{d}^t||_2^2}{
  \label{awsr_optimization}}{}
  \addConstraint{\sum_{k'\in\mathcal{K}}\bar{C}_{k'}}{\leq\bar{R}_{c,k}(\hat{\mathbf{P}})\;,\;\forall k\in\mathcal{K}}
  \addConstraint{\bar{\mathbf{c}}}{\geq \mathbf{0}}
  \addConstraint{\diag(\hat{\mathbf{P}}\hat{\mathbf{P}}^H)}{=\frac{P_t\mathbf{1}}{N_t}}
  \addConstraint{(\bar{C}_k + \bar{R}_k(\hat{\mathbf{P}}))}{\geq \bar{R}_k^{th}\;,\;\forall k\in\mathcal{K}}.
\end{mini}

Due to partial CSIT, the problem in (\ref{awsr_optimization}) is stochastic in nature. To solve it, the method proposed in \cite{joudeh} is adapted. Therefore, (\ref{awsr_optimization}) is first converted into a deterministic problem by employing the Sampled Average Approximation (SAA) method. Then, it is further transformed into a convex problem by applying the Weighted Minimum Mean Squared Error (WMMSE) approach and solved by using the Alternating Optimization (AO) algorithm.

\subsection{BSE Minimization Sub-problem}
The $\mathbf{u}$-update sub-problem in (\ref{uupdate}) is reformulated as follows:
\begin{mini}|s|[2]
  {\alpha_u,\mathbf{p}_u}{\lambda\sum_{m=1}^M|\alpha_u \bm{P}_d(\theta_m)-\mathbf{a}^H(\theta_m)\big(\sum_{k=1}^{K+1}\mathbf{D}_{p,k}\mathbf{p}_u\mathbf{p}_u^H\mathbf{D}_{p,k}^H\big)}{
  \label{radar_optimization}}{}
  \breakObjective{\mathbf{a}(\theta_m)|^2+\frac{\rho}{2}||\mathbf{D}_p\mathbf{v}^{t+1}-\mathbf{p}_u+\mathbf{d}^t||_2^2}
  \addConstraint{\diag\big(\sum_{k=1}^{K+1}\mathbf{D}_{p,k}\mathbf{p}_u\mathbf{p}_u^H\mathbf{D}_{p,k}^H\big)}{=\frac{P_t\mathbf{1}}{N_t}}
  \addConstraint{\alpha_u}{>0},
\end{mini}
where $\alpha_u=u_1$ is the first entry of the optimization variable $\mathbf{u}$, $\mathbf{p}_u=[u_{K+2},u_{K+3},\dots,u_{(N_t+1)\times(K+1)}]^T \in \mathbb{C}^{N_t(k+1) \times 1}$, and $\mathbf{D}_{p,k} = [\mathbf{0}^{N_t \times (k-1)N_t}, \mathbf{I}_{N_t}, \mathbf{0}^{N_t \times (K+1-k)N_t}]$. Although ($\ref{radar_optimization}$) is originally non-convex, it can be changed into a convex expression by employing Semi-Definite Relaxation (SDR) techniques \cite{sdr}. 

\subsection{ADMM Algorithm}
The ADMM-based optimization algorithm is summarized in Algorithm $\ref{radcom_admm_algorithm_imperfect}$. The process is repeated iteratively until the primal residual $\mathbf{r}^{t+1}$ and the dual residual $\mathbf{q}^{t+1}$ of the ADMM algorithm converge to a value below a predefined threshold $\nu$.

\begin{algorithm}
\DontPrintSemicolon
  \KwInput{$t \leftarrow 0$, $\mathbf{v}^t$, $\mathbf{u}^t$,
  $\mathbf{d}^t$;}
  \Repeat{$||\mathbf{r}^{t+1}||_2 \leq \nu$ and $||\mathbf{q}^{t+1}||_2 \leq \nu$}
  {
  $\mathbf{v}^{t+1}\leftarrow\arg \min_{\mathbf{v}}\big(f_c(\mathbf{v}) + g_c(\mathbf{v})+(\rho/2)||\mathbf{D}_p(\mathbf{v}-\mathbf{u}^t)+\mathbf{d}^t||_2^2\big) $ using SAA AR-WMMSE-AO;\\
  $\mathbf{u}^{t+1}\leftarrow\arg \min_{\mathbf{u}}\big(f(\mathbf{u}) + g( \mathbf{u})+(\rho/2)||\mathbf{D}_p(\mathbf{v}^{t+1}-\mathbf{u})+\mathbf{d}^t||_2^2\big)$ using SDR;\\
  $\mathbf{d}^{t+1}\leftarrow\mathbf{d}^t+\mathbf{D}_p(\mathbf{v}^{t+1}-\mathbf{u}^{t+1});$\\
  $\mathbf{r}^{t+1}=\mathbf{D}_p(\mathbf{v}^{t+1}-\mathbf{u}^{t+1});$\\
  $\mathbf{q}^{t+1}=\mathbf{D}_p(\mathbf{u}^{t+1}-\mathbf{u}^{t});$\\
  $t \leftarrow t+1$; 
  }
\caption{ADMM-based RadCom optimization algorithm with partial CSIT}
\label{radcom_admm_algorithm_imperfect}
\end{algorithm}

\section{Performance Evaluation}
\label{numerical_section}
In this section, the joint RSMA-RadCom is evaluated in terms of its Ergodic Weighted Sum-Rate (EWSR) and Ergodic Root Beampattern Squared Error (ERBSE) trade-off, where the average of the optimization results for 200 different channel realizations are used. It is assumed that the radar target is located at the 0° azimuth direction, $P_t = 20$ dBm, $N_t=4$, $\delta=0.5$, $K=2$, $\mu_k = 1/K\;\forall\; k\in \mathcal{K}$, $\rho=1$, $\nu = 10^{-2}$, and the QoS rate constraint $\bar{R}_k^{th}=1$ bps/Hz, $\;\forall\; k\in \mathcal{K}$. Also, $\sigma_k^2=1,\forall k\in \mathcal{K}$ is used to generate the user channel vectors, the CSIT quality scaling factor for all $K$ users is $\alpha = 0.6$, and $\bm{\lambda}=[10^{-9},10^{-8},\dots,10^{-1}]^T$ is the regularization parameter vector, where increasing lambda shifts the priority from communications to radar sensing. In order to highlight the gains brought by empowering the RadCom with RSMA, the use of Space Division Multiple Access (SDMA) is considered. SDMA fully treats MUI as noise, so SDMA operation is enabled by not allocating any power to the common stream precoder in (\ref{rsma_transmit_signal}) and omitting $\bar{C}_k$ in (\ref{awsr_def}) \cite{rsma_lina}. Results for perfect CSIT optimization as described in \cite{chengcheng} are also included in order to demonstrate the robustness of the RSMA-RadCom as the CSIT quality degrades. 
\begin{figure*}[t]
\centering
\begin{minipage}[b]{.47\textwidth}
   \centering
   \centering
        \includegraphics[scale = 0.64]{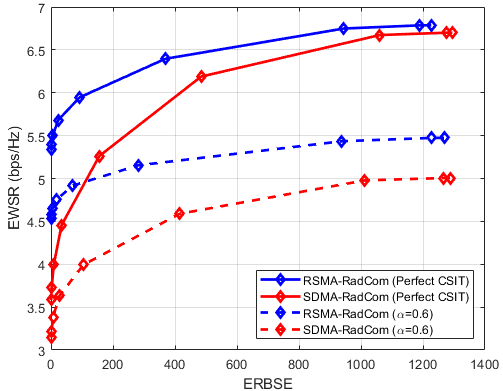}
		\caption{EWSR - ERBSE trade-off.}
		\label{fig:EWSR_ERBSE}
\end{minipage}\qquad
\begin{minipage}[b]{.47\textwidth}
   \centering
       \includegraphics[scale = 0.40]{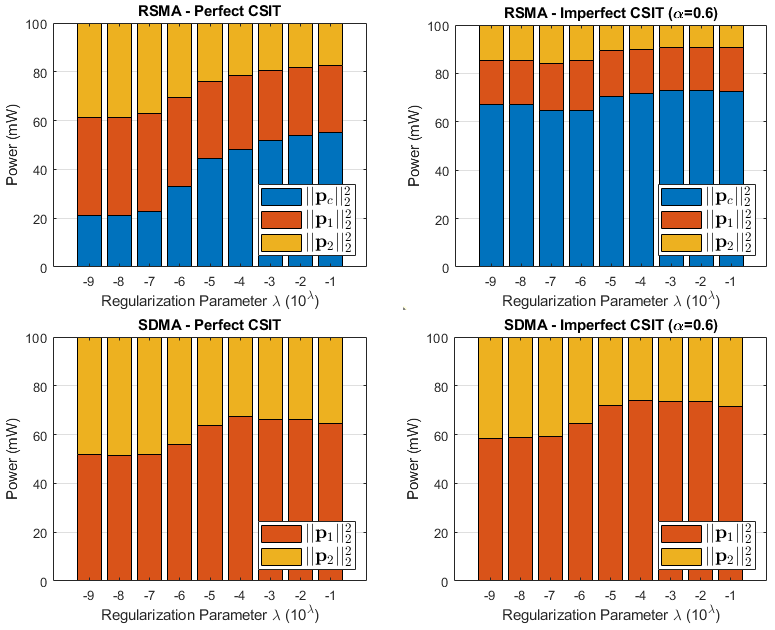}
		\caption{Ergodic Precoder Power Allocation.}
		\label{fig:ergodicPow}
\end{minipage}
\end{figure*}

The generated ergodic trade-off curves are plotted in Fig. \ref{fig:EWSR_ERBSE} and the ergodic precoder power allocation is shown in Fig. \ref{fig:ergodicPow}. With Perfect CSIT, the RSMA-RadCom and SDMA-RadCom show similar EWSR levels when communications are prioritized. This is the effect of directly maximizing the instantaneous user rates  by mainly employing the user private streams as observed in Fig. \ref{fig:ergodicPow}. Nonetheless, the RSMA-RadCom still presents a slightly better trade-off by employing its common stream to jointly mitigate the MUI and approximate the desired radar beampattern in cases where the communication users are located in azimuth directions near the radar target. As Radar is given more priority, it is observed that the ERBSEs of the RadComs become more identical but the EWSR of the SDMA-RadCom decays at a faster. This can be explained by noticing from Fig. \ref{fig:ergodicPow} that the RSMA-RadCom starts allocating more power to the common stream to generate a directional beampattern, which also assists in not increasing the MUI for user-2 in the same level as the SDMA-RadCom. For full radar priority, both RadComs achieve ERBSE = 0 but the EWSR of the RSMA-RadCom is 1.75 bps/Hz larger than that of the SDMA-RadCom. 

With partial CSIT, it is observed that the RSMA-RadCom still outperforms the SDMA-RadCom and when contrasting the trade-off curves with their Perfect CSIT counterparts, it is seen that each of them is affected to a different degree. For instance, the RSMA-RadCom now outperforms the SDMA-RadCom by 0.47 bps/Hz for communications, and by 1.34 bps/Hz for radar. From Fig. \ref{fig:ergodicPow}, it can be noticed that the RSMA-RadCom now employs the common stream to a larger degree to combat MUI imposed by CSIT estimation errors and by also forcing the MIMO radar beampattern generation. In turn, the SDMA-RadCom has no other option but to allocate more power to the precoder of user-1 to maximize the AWSR and to generate the MIMO radar beampattern. This inevitably increases the MUI to user-2 and, hence, it achieves a much lower ergodic rate compared to user-1. 

\section{Conclusion}
\label{conclusion_section}
An ADMM-based algorithm is introduced which optimizes the precoders of a joint RadCom to simultaneously maximize the AWSR and minimize the BSE against a desired highly-directional transmit beampattern in the presence of partial CSIT. Analysis of the ergodic performance of the RadCom reveals that a RSMA-aided approach enables a more robust and flexible joint operation to comply with QoS rate constraints than SDMA. These benefits are due to mainly employing the common stream to mitigate the MUI introduced by CSIT inaccuracies and to approximate a directional MIMO radar beampattern.

\ifCLASSOPTIONcaptionsoff
  \newpage
\fi

\end{document}